\documentclass[12pt a4paper]{article}

\usepackage{latexsym}
\usepackage{graphicx}
\usepackage{amsmath}
\usepackage{amssymb}
\graphicspath{{./}{./figs/}}

\begin{document}

\title{\bf Uniform synchronous criticality of diversely random complex networks}
\author{{\bf Xiang Li\thanks{E-mail: xli@sjtu.edu.cn, Fax: +86-21-62932344}}\\
Department of Automation\\
Shanghai Jiao Tong University\\
Shanghai, 200030, P.R.China\\
\\Accepted by Physica A\\}

\maketitle

\begin{abstract}
We investigate collective synchronous behaviors in random complex
networks of limit-cycle oscillators with the non-identical
asymmetric coupling scheme, and find a uniform coupling
criticality of collective synchronization which is independent of
complexity of network topologies. Numerically simulations on
categories of random complex networks have verified this
conclusion.\\


\end{abstract}

\section{Introduction}
Traditionally, a network with complex topology is mathematically
described by a random graph from the ER model proposed by
Erd\"{o}s and R\'enyi \cite{E-R:1960}, which has ever dominated
scientists' thinking about complex networks for almost 40 years.
Till recently, the discoveries of small-world networks \cite
{W-S:1998} and scale-free networks \cite{B-A:1999} pioneered a
revolution in the theoretic study of complex networks, and,
surprisingly, many real-life large-scale complex networking
systems are in the categories of small-world and(or) scale-free
networks, ranging from biological to engineering systems, and even
to social and economic systems \cite{A-B:2002}-\cite{L-J-C}.

It has been argued that network topologies significantly affect
the emergent network dynamics \cite{Strogatz:2001}, and the
small-world phenomena and scale-free features of complex networks
have led to a fascinating set of common problems concerning how
the complexity of network topology facilitates and constrains
synchronous behaviors of networks \cite{Wang:2002}.

Collective synchronization in a large population of oscillators
having natural different frequencies is one of the focal points in
the synchronization literature, which is a typical phenomenon in
the fields of biology, physics, and engineering
\cite{Strogatz:2000}. After the first mathematical description by
Wiener \cite{Wiener:1961} and the later fruitful study by Winfree
\cite{Winfree:1967}, Kuramoto refined this connection between
nonlinear dynamics and statical physics, and formalized the
solution to a network of globally coupled limit-cycle oscillators
\cite{Kuramoto:1984}, answering the situation why the oscillators
are completely de-synchronized until the coupling strength
overcomes a criticality $C_{syn}$.

The recent discoveries of topological complexity accelerate the
understanding of this cooperative phenomenon in the frame of
random complex networks \cite{H-C-K:2002}-\cite{M-V-P:2004}.
Especially, Hong et al. reported their synchronization
observations of a larger criticality of coupling strength on
small-world networks than that of globally coupled networks
\cite{H-C-K:2002}. And, the latest investigation stated the
absence of critical coupling strength in frequency synchronization
of a swarm of oscillators connected as a scale-free network having
a power-law exponent $2<\gamma\le 3$ \cite{Ichinomiya:2004}.

All these fruits indicate that the topology of complex networks
does play an important role in collective synchronization, and
different categories of networks may hold significantly different
critical coupling strengths for the occurrence of synchrony in
mutually coupled oscillators. However, it should be pointed out
that most of these issues hold an assumption that every pair of
connected nodes are coupled together with the identical coupling
strength. In practice, it is more general that pairs of nodes are
connected with non-identical and asymmetric couplings. Therefore,
it is natural to ask the question that whether the collective
synchronous behaviors of complex networks with non-identical and
asymmetric coupling schemes still depend on the network
topologies? In this paper, we try to explore this question, and
the answer is that there does exist a uniform criticality of
coupling strength to synchronize diversely random networks of
limit-cycle oscillators.

The whole paper is organized as follows. Section II describes the
phase evolving model of a complex network of limit-cycle
oscillators, whose couplings are non-identical and asymmetric as
specified in this work. A uniform critical coupling strength for
collective synchrony among oscillators is analytically arrived at
the same section, whose validity is verified for categories of
random complex networks by numerical simulations in Section III.
Section IV finally concludes this investigation.

\section{Model description and main result}
We consider a network of $N$ coupled limit-cycle oscillators whose
phases $\theta_i, i=1,2,\cdots,N,$ evolve as
\begin{equation}
\label{eq0a}\frac{d\theta_i}{dt}=\omega_i+\sum_{j=1}^N
C_{ij}a_{ij}\mbox{sin}(\theta_j-\theta_i),
\end{equation}
where $C_{ij}$ is the coupling strength between node (oscillator)
$i$ and node (oscillator) $j$, and $a_{ij}$ is 1(or 0) if node $i$
is connected (or disconnected) with node $j$. Frequencies
$\omega_i, i=1,2,\cdots, N,$ are randomly distributed following
the given frequency distribution $g(\omega)$, which is assumed
that $g(\omega)=g(-\omega)$.

Define the non-identical asymmetric coupling scheme
\begin{equation}
\label{eq0b} C_{ij}=C_j=\frac{C}{k_i},\qquad i,j=1,\cdots,N,
\end{equation}
where $C$ is a positive constant, and $k_i$ is the degree of node
$i$, which fits the given degree distribution $P(k)$ of a network.
Therefore, we have
\begin{equation}
\label{eq1}\frac{d\theta_i}{dt}=\omega_i+\frac{C}{k_i}\sum_{j=1}^N
a_{ij}\mbox{sin}(\theta_j-\theta_i).
\end{equation}
If for every node $i$, its degree $k_i=N, i=1,2,\cdots, N$, model
(\ref{eq1}) equals the classic Kuramoto model for globally coupled
networks in \cite{Kuramoto:1984,Strogatz:2001,Strogatz:2000}.

Define the order parameter $(r,\Psi)$ as \cite{Ichinomiya:2004}:
\begin{equation}
\label{eq2}re^{i\Psi}=\frac{\int d\omega\int dk \int d\theta
g(\omega)P(k)k\rho(k,\omega;t,\theta)e^{i\theta}}{\int dkP(k)k},
\end{equation}
where $\rho(k,\omega;t,\theta)$ is the density of oscillators with
phase $\theta$ at time $t$ with the given frequency $\omega$ and
degree $k$, which satisfies the normalization as
\begin{equation}
\label{eq3} \int_0^{2\pi}\rho(k,\omega;t,\theta)d\theta=1.
\end{equation}

Assume $v$ to be the continuum limit of the right-hand side
(r.h.s.) of Eq. (\ref{eq1}), and each randomly selected edge is
connected to the oscillator having degree $k$, frequency $\omega$,
and phase $\theta$ with probability
$\frac{kP(k)g(\omega)\rho(k,\omega;t,\theta)}{\int dkP(k)k}$.
Therefore, determined by the continuity equation
\begin{equation}
\label{eq4}\frac{\partial\rho}{\partial t}=-\frac{\partial (\rho
v)}{\partial\theta},
\end{equation}
the density $\rho(k,\omega;t,\theta)$ evolves as
\begin{equation}
\label{eq5} \frac{\partial \rho(k,\omega;t,\theta)}{\partial
t}=-\frac{\partial}{\partial \theta}\left
[\rho(k,\omega;t,\theta)\left (\omega+\frac{C}{k}k\frac{\int
d\omega'\int dk'\int d\theta'
g(\omega')P(k')k'\rho(k',\omega';t,\theta')}{\int
dk'P(k')k'}\mbox{sin}(\theta-\theta')\right ) \right ].
\end{equation}

Substituting Eq. (\ref{eq2}) into Eq. (\ref{eq5}) yields
\begin{equation}
\label{eq6}\frac{\partial \rho (k,\omega;t,\theta)}{\partial
t}=-\frac{\partial}{\partial \theta}\left
\{\rho(k,\omega;t,\theta)\left
[\omega+Cr\mbox{sin}(\Psi-\theta)\right ]\right \},
\end{equation}
whose solution independent of time is
\begin{equation}
\label{eq7}\frac{\partial}{\partial \theta}\left
\{\rho(k,\omega;\theta)\left
[\omega+Cr\mbox{sin}(\Psi-\theta)\right ]\right \}=0.
\end{equation}
where $\rho(k,\omega;\theta)$ could be assumed to be
\begin{equation}
\label{eq8} \rho(k,\omega;\theta)=\left \{\begin{array}{l}
\delta\left (\theta-\mbox{arcsin}\left (\frac{\omega}{Cr}\right )
\right )\;\;\;\;\mbox {if}\;\frac{|\omega|}{Cr}\le 1\\
\frac{D(k,\omega)}{|\omega-Cr\mbox{sin}\theta|}\;\;\;\;\qquad\qquad\mbox{otherwise.}
\end{array}\right .
\end{equation}
Here, we assume $\psi=0$ without loss of generality, and
$D(k,\omega)$ is the normalized factor. Recall
$g(-\omega)=g(\omega)$, and substitute Eq. (\ref{eq8}) into Eq.
(\ref{eq2}). We therefore arrive at
\begin{eqnarray}
\label{eq9}r=\frac{\int dk\int_{-Cr}^{Cr} d\omega
g(\omega)kP(k)e^{i\;\mbox{arcsin}\;\left(\frac{\omega}{Cr}\right)}}{\int
dkkP(k)}\\
=\int_{-Cr}^{Cr}d\omega
g(\omega)\sqrt{1-\left(\frac{\omega}{Cr}\right)^2}\qquad\qquad\quad\\
=Cr\int_{-\frac{\pi}{2}}^{\frac{\pi}{2}}\mbox{cos}^2\theta
g(Cr\mbox{sin}\theta)d\theta.\qquad\quad\quad\quad\;
\end{eqnarray}
From this self-consistency equation we finally come to
\begin{equation}
\label{eq10} C_{syn}=\frac{2}{\pi g(0)},
\end{equation}
which is surprisingly the same as the critical coupling strength
of globally coupled networks investigated by Kuramoto
\cite{Kuramoto:1984,Strogatz:2000}.

The main point of criticality (\ref{eq10}) is that, with the
non-identical and asymmetric coupling scheme (\ref{eq0b}),
different networks may exhibit the same transition of collective
synchronous behaviors regardless of the significant difference of
their degree distributions, i.e., the complexity of network
topology. We will verify this point through extensive simulations.

\section{Numerical simulations}
Define the average order parameter as
\begin{equation}
\label{eq11}
r_{av}=\left<\left[\frac{\sum_{i=1}^Nk_ie^{i\theta_i}}{\sum_{i=1}^N
k_i}\right]\right>,
\end{equation}
where $\left<\cdots\right>$ and $[\cdots]$ denote the averages
over different realizations of intrinsic networks and over
different realizations of intrinsic frequencies, respectively. In
all the simulations, $r_{av}$ is averaged over 10 groups of
networks satisfying the same degree distribution $P(k)$, and each
network has 5 sets of frequencies with the distribution
$g(\omega)$. We further specify
\begin{equation}
\label{eq12} g(\omega)=\left\{\begin{array}{l}
0.5,\qquad\mbox{if} -1<\omega<1\\
0,\qquad\quad\mbox{otherwise}.\end{array}\right .
\end{equation}
Therefore, $C_{syn}\approx 1.273$ by Eq.(\ref{eq10}) and Eq.
(\ref{eq12}).

\begin{figure}
\includegraphics[width=12cm]{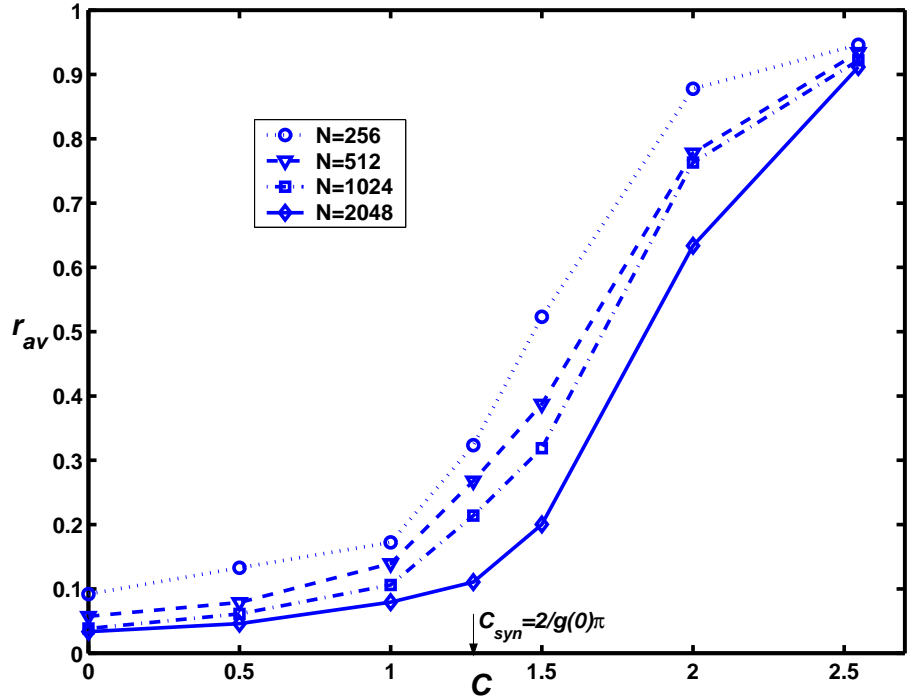}
\caption{\label{fig-KuraBARav} The average order parameter
$r_{av}$ vs the coupling strength $C$ for scale-free networks of
the BA model with $N=256,512,1024,2048$, respectively. All
networks are started from $m=m_0=3$.}
\end{figure}

\begin{figure}
\includegraphics[width=12cm]{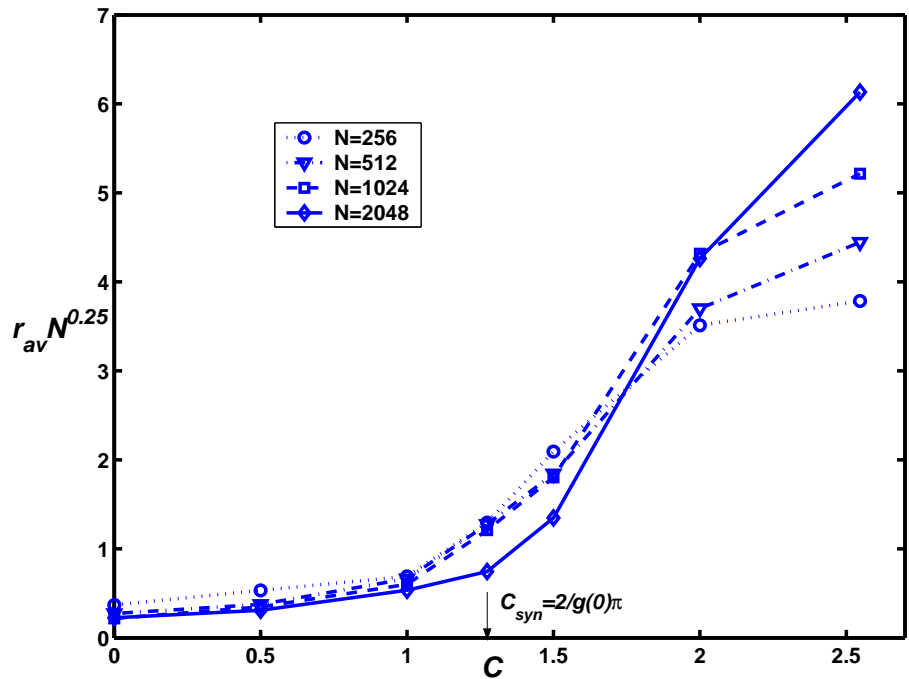}
\caption{\label{fig-KuraBARavN} The average $r_{av}N^{0.25}$ vs
the coupling strength $C$ for scale-free networks of the BA model
with $N=256,512,1024,2048$, respectively. All networks are started
from $m=m_0=3$.}
\end{figure}

We first numerically simulate the case of scale-free networks of
the BA model \cite{B-A-J:1999} whose degree distribution is in the
power-law form of $P(k)\propto k^{-3}$. Figures
\ref{fig-KuraBARav}-\ref{fig-KuraBARavN} show the simulating
outcomes on scale-free networks of BA model with $m=m_0=3$, and
$N=256,512,1024,2048$, respectively. It can be clearly observed
that even having the finite-size effect for $N=256,512,1024,2048$,
respectively, the average order parameter $r_{av}$ shows a rapid
increase after the critical coupling strength
$C_{syn}=\frac{2}{g(0)\pi}$, indicating that when $C_{syn}\le C$,
the scale-free network of oscillators begin to cluster with
synchrony. In Fig. \ref{fig-KuraBARavN}, $r_{av}N^{0.25}$ shows a
similar dependence on the coupling strength $C$, which rapidly
increased when $C_{syn}\le C$.

\begin{figure}[t]
\includegraphics[width=12cm]{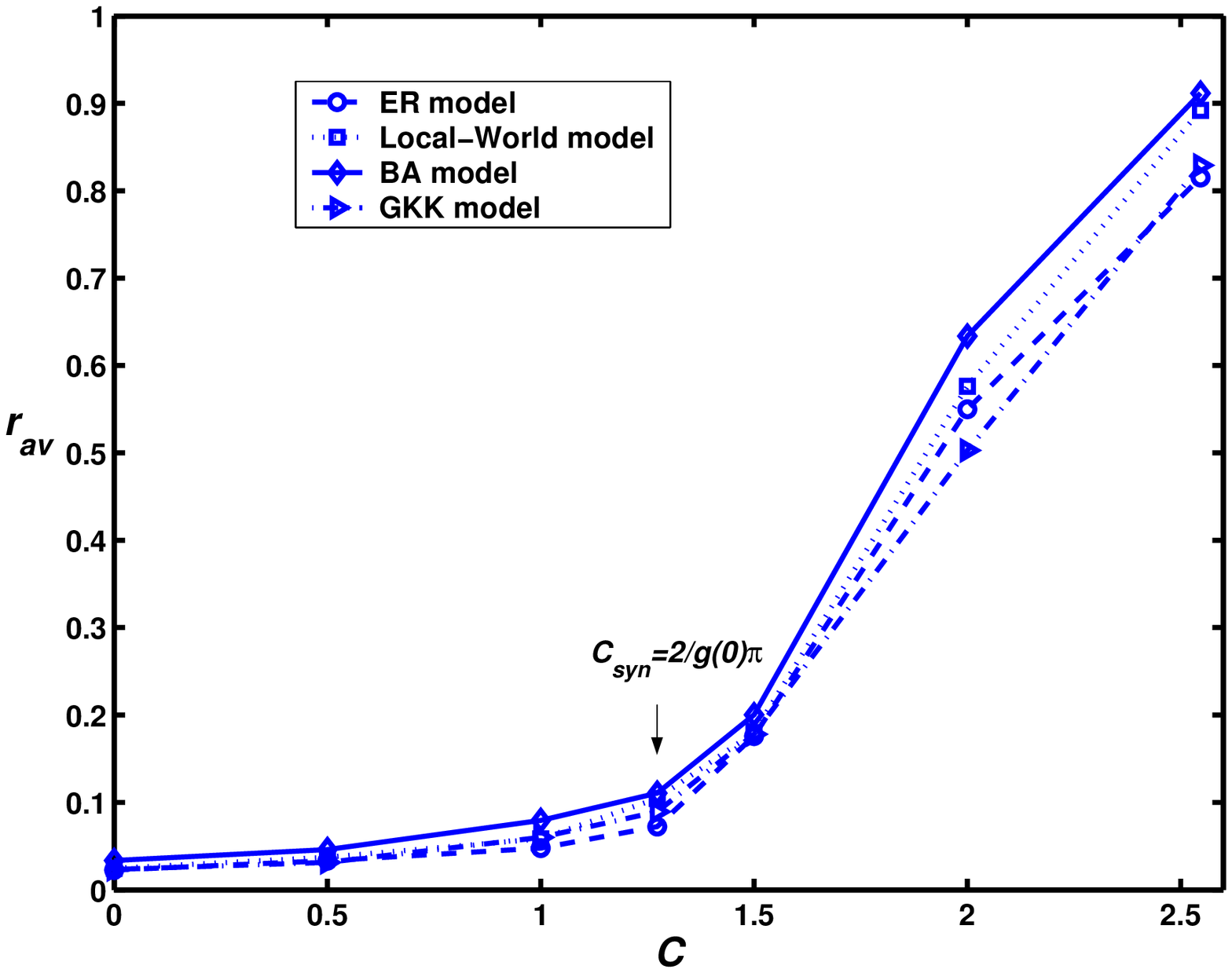}
\caption{\label{fig-KuraNet2048Rav} The average order parameter
$r_{av}$ vs the coupling strength $C$ for networks generated by
the ER model (dashed line with circle markers), the Local-World
model (dotted line with square markers), the BA model (solid line
with diamond markers), and the GKK model having power-law exponent
$\gamma=6$ (dash-dot line with right triangle markers). All
networks have the same scale $N=2048$ and the same average degree
$\left<k\right>=6$.}
\end{figure}

\begin{figure}[t]
\includegraphics[width=12cm]{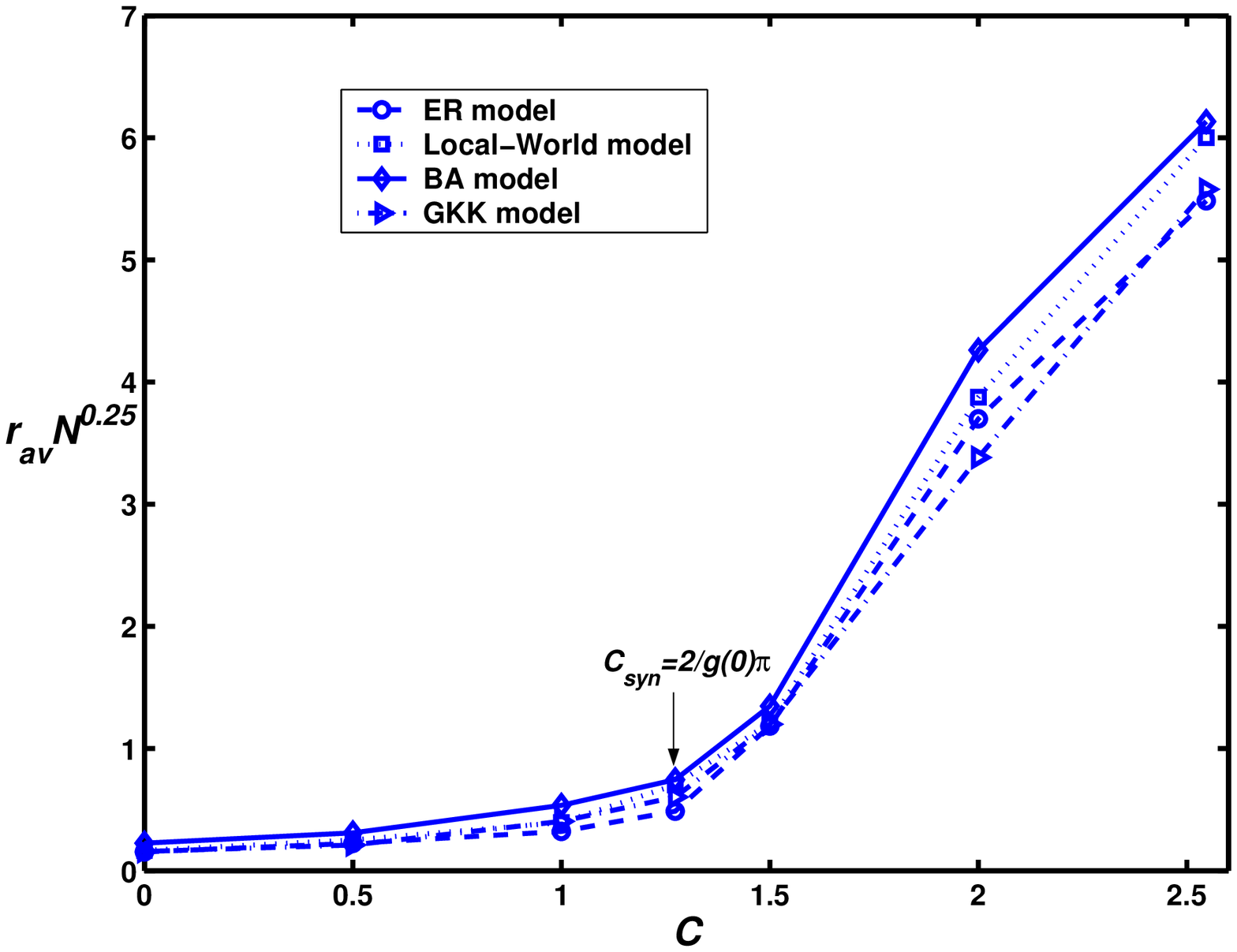}
\caption{\label{fig-KuraNet2048RavN} The average $r_{av}N^{0.25}$
vs the coupling strength $C$ for networks generated by the ER
model (dashed line with circle markers), the Local-World model
(dotted line with square markers), the BA model (solid line with
diamond markers), and the GKK model having power-law exponent
$\gamma=6$ (dash-dot line with right triangle markers). All
networks have the same scale $N=2048$ and the same average degree
$\left<k\right>=6$.}
\end{figure}

From this simulation example of the same scale-invariant power-law
degree distribution $P(k)\propto k^{-3}$ with different network
scale $N$, we clearly conclude that the nonzero constant
criticality $C_{syn}$ is independent of the finite size of a
network, showing a difference from the investigated case of
scale-free identical-coupled oscillators in
\cite{Ichinomiya:2004}.

To investigate the independence of $C_{syn}$ on different degree
distributions $P(k)$, we fix the scale $N=2048$, and the average
degree $\left<k\right>=6$. Therefore, all networks have the same
number of nodes and edges with different connectivity patterns in
the next simulation studies.

Two main categories of degree distributions of random complex
networks are in the forms of $P_{power}(k)\propto k^{-\gamma}$ and
$P_{exp}(k)\propto e^{-k}$. Therefore, we select the famous ER
model \cite{E-R:1960} to generate networks having an exponential
degree distribution $P_{exp}(k)\propto e^{-k}$, and nominate the
BA model and the GKK model \cite{G-K-K:2001} to generate networks
having the scale-invariant power-law degree distribution
$P_{power}(k)\propto k^{-\gamma}$ with $\gamma=3,6$, respectively.
The proposed local-world evolving network \cite{L-C:2003} owns a
transition between the exponential degree distribution and the
power-law degree distribution, therefore it is adopted as the
final prototype for the coming simulations with the parameters
$M=10$, $m=m_0=3$.

The simulating results of Figs.
\ref{fig-KuraNet2048Rav}-\ref{fig-KuraNet2048RavN} are not out of
expectation of criticality (\ref{eq10}). There is a common
critical coupling strength of the value
$C_{syn}=\frac{2}{g(0)\pi}$ in these four categories of networked
limit-cycle oscillators, and both the average order parameter
$r_{av}$ and the average $r_{av}N^{0.25}$ increased sharply when
$C_{syn}\le C$. In other words, the significant difference among
complex network topologies does not show effect on the collective
synchrony of the non-identically asymmetrically coupled
limit-cycle oscillators (\ref{eq1}). However, as concluded from
\cite{H-C-K:2002,M-V-P:2004,Ichinomiya:2004}, there is no such a
same critical coupling strength of collective synchrony in
different random complex networks of identically symmetrically
coupled limit-cycle oscillators.

\section{Conclusion}
To summarize, we have investigated the collective synchronous
behaviors of complex networks of oscillators having non-identical
asymmetric coupling strengths, which show a uniform synchrony
criticality regardless of the complexity of networking topologies.
One of the direct extensions is that for scale-free networks, when
synchronization is not preferred in some practical situations
\cite{L-C:2003-2}, the proposed asymmetric coupling scheme
(\ref{eq0b}) could be applied to obtain a nonzero synchrony
criticality which is independent of the network topologies. More
important to note is the investigation in this paper indicates
that, with the embedment of complicated coupling schemes, the
interactions previously observed between the network collective
behaviors and network complex topologies are still a small tip of
an huge iceberg, most of which are left for further explorations.

\section*{Acknowledgements}
The author thanks Dr. Takashi Ichinomiya for his valuable
discussion. This work was partially supported by the National
Natural Science Foundation of P.R. China under Grants No. 90412004
and 70431002, 
and the author also acknowledges the support from the Alexander
von Humboldt Foundation and the SRF for ROCS, SEM.


\begin{thebibliography}{99}

\bibitem{E-R:1960}
P. Erd\"{o}s and A. R\'enyi, Publicationes Mathematicae {\bf 6},
290 (1959); Publ. Math. Inst. Hung. Acad. Sci. {\bf 5}, 17 (1960);
Acta Mathematica Scientia Hungary, {\bf 12}, 261 (1961).

\bibitem{W-S:1998}
D.J. Watts and S.H. Strogatz, Nature {\bf 393}, 440 (1998).

\bibitem{B-A:1999}
A.L. Barab$\acute{a}$si and R. Albert, Science {\bf 286}, 509
(1999).

\bibitem{A-B:2002}
R. Albert and A.L. Barab$\acute{a}$si, Reviews of Modern Physics
{\bf 74}, 47 (2002).

\bibitem{D-M:2002}
S.N. Dorogovstsev and J.F.F. Mendes, Advances in Physics {\bf 51},
1079 (2002).

\bibitem{Newman:2003}
M.E.J. Newman, SIAM Review {\bf 45}, 167 (2003).

\bibitem{W-C:2003}
X.F. Wang and G. Chen, IEEE Circuits and Systems Magazine {\bf 3},
6 (2003).

\bibitem{L-J-C}
X. Li, Y.Y. Jin and G. Chen, Physica A {\bf 328}, 287 (2003);
Physica A {\bf 343}, 573 (2004).

\bibitem{Strogatz:2001}
S.H. Strogatz, Nature {\bf 410}, 268 (2001).

\bibitem{Wang:2002}
X.F. Wang, International Journal of Bifurcation and Chaos {\bf
12}, 885 (2002).

\bibitem{Strogatz:2000}
S.H. Strogatz, Phyisca D {\bf 143}, 1 (2000).

\bibitem{Wiener:1961}
N. Wiener, {\sl Cybernetics} (MIT Press, Cambridge, MA, 1961).

\bibitem{Winfree:1967}
A.T. Winfree, J. Theoret. Biol. {\bf 16}, 15 (1967).

\bibitem{Kuramoto:1984}
Y. Kuramoto, {\sl Chemical Oscillations, Waves, and Turbulence}
(Springer, Berlin, 1984).

\bibitem{H-C-K:2002}
H. Hong, M.Y. Choi, and B.J. Kim, Phys. Rev. E {\bf 65}, 026139
(2002).

\bibitem{Ichinomiya:2004}
T. Ichinomiya, Phys. Rev. E {\bf 70}, 026116 (2004).

\bibitem{M-V-P:2004}
Y. Moreno, M. V$\acute{a}$zquez-Prada, and A.F. Pacheco, Physica A
{\bf 343}, 279 (2004); Y. Moreno and A.F. Pacheco,
cond-mat/0401266 (2004); D.S. Lee, cond-mat/0410635 (2004).

\bibitem{B-A-J:1999}
A.L. Barab$\acute{a}$si, R. Albert, and H. Jeong, Physica A {\bf
272}, 173 (1999).

\bibitem{L-C:2003}
X. Li and G. Chen, Physica A {\bf 328}, 274 (2003).

\bibitem{G-K-K:2001}
K.I. Goh, B. Kahng, and D. Kim, Phys. Rev. Lett. {\bf 87}, 278701
(2001).

\bibitem{L-C:2003-2}
X. Li and G. Chen, IEEE Trans. Circuits and Systems-I {\bf 50},
1381-1390 (2003).

\end{thebibliography}
\end{document}